\documentclass[12pt,preprint]{aastex}
\usepackage{natbib}
\usepackage{graphicx}
\usepackage{amssymb}
\usepackage{amsmath}
\shorttitle{The $d^*$(2380) hexaquark Bose Einstein Condensate dark matter}
\begin{document}
\title{The decaying and scattering properties of the $d^*$(2380) hexaquark Bose Einstein Condensate dark matter}
\author{Man Ho Chan}
\affil{Department of Science and Environmental Studies, The Education University of Hong Kong, Hong Kong, China}
\email{chanmh@eduhk.hk}

\begin{abstract}
Recently, a study has shown that the Bose Einstein Condensates formed by the $d^*$(2380) hexaquarks ($d^*$(2380)-BECs) can be thermally produced in the early universe and they are stable enough to be a competitive candidate of dark matter. Searching for the decaying signature of $d^*$(2380)-BECs is a possible way to verify this dark matter model. In this article, we discuss the scattering and decaying properties of the $d^*$(2380)-BECs and we show that the decay rate of the $d^*$(2380)-BECs is correlated with the TeV cosmic-ray flux. The predicted average decay rate in our Galaxy is several orders of magnitude larger than the current observed upper limit. Therefore, it would be very difficult for us to search for the decaying signature of the $d^*$(2380)-BEC dark matter model. Nevertheless, the size of the $d^*$(2380)-BECs may be large enough to have self-interaction so that we can possibly detect them in the future.
\end{abstract}


\section{Introduction}
Observational data of galaxies, galaxy clusters and the cosmic microwave background reveal that some unknown dark matter particles exist in our universe. However, all of the known fundamental particles in the Standard Model do not exhibit the properties of dark matter. Although many theoretical models have suggested some possible dark matter candidates such as Weakly Interacting Massive Particles (WIMPs) or sterile neutrinos, there is no promising observed signal of these hypothetical particles so far. Current observational data of direct detections \citep{Tan,Aprile,Aprile2}, indirect detections (gamma-ray, radio or cosmic-ray detections) \citep{Calore,Daylan,Abazajian,Ackermann,Albert,Chan,Egorov,Chan2,Chan3,Chan4,Boudaud,Bergstrom,Cavasonza,Aguilar} and collider experiments \citep{Abecrcrombie} have ruled out a large parameter space of particle dark matter models, especially for WIMPs \citep{Roszkowski}.

Many previous models of particle dark matter assume that they are fermions (e.g. WIMP models). Nevertheless, recently, many studies are now focusing on dark matter particles being bosons. One important feature of bosonic dark matter is that the bosonic dark matter particles can form a Bose Einstein Condensate (BEC) while femionic dark matter cannot \citep{Chavanis}. For example, if the mass of the bosonic dark matter particles is $m \sim 10^{-22}$ eV, then they can form a very large BEC and behave like a large dark matter halo in a galaxy or galaxy cluster \citep{Zhang}. 

Recently, a study has shown that a certain number of hexaquarks $d^*$(2380) can be bounded together to form a stable BEC (hereafter called $d^*$(2380)-BEC) \citep{Bashkanov}. The $d^*$(2380)-BECs can be thermally formed in the early universe \citep{Bashkanov}. The hexaquark $d^*$(2380) is formed by six quarks (3 $u$ quarks and 3 $d$ quarks) and its existence was confirmed in collider experiments in the past decade \citep{Adlarson,Adlarson2,Lu}. The mass of an individual $d^*$(2380) hexaquark is $m_d=2.38$ GeV while the mass of a $d^*$(2380)-BEC can be larger than 1 TeV, which depends on the total number of bounded hexaquarks.

A $d^*$(2380)-BEC could break down and decay to emit gamma rays with energy $\sim 100-500$ MeV \citep{Bashkanov}. In this article, we theoretically discuss the scattering properties and the decay rate $\Gamma_d$ of the $d^*$(2380)-BECs. We show that the observed value of $\Gamma_d$ may not be a constant and it depends on the astrophysical environment. We also constrain the average $\Gamma_d$ in our Galaxy and compare it with our theoretical prediction.      

\section{Theoretical prediction of the scattering rate and decay rate}
A group of $d^*$(2380) hexaquarks can form stable BECs. The binding energy depends on the number of hexaquarks bounded and the geometrical shapes (e.g. spherical shape) of the BECs. The binding energy $B$ of the $d^*$(2380)-BEC per number of hexaquarks $D$ is given by \citep{Bashkanov}
\begin{equation}
\frac{B}{D}=a_V(D-1)-a_C \frac{D}{D^{1/3}},
\end{equation}
where $a_V \sim 1$ MeV and $a_C \sim 0.1$ MeV are coefficients that determine the relative strengths of the attractive volume and the repulsive Coulomb terms. A stable $d^*$(2380)-BEC may consist of $\sim 10^3-10^6$ $d^*$(2380) hexaquarks. The minimum $D$ for a stable $d^*$(2380)-BEC is $D \sim 10^3$ so that the binding energy threshold of a stable $d^*$(2380)-BEC is $\sim 1$ TeV \citep{Bashkanov}. A large amount of stable $d^*$(2380)-BECs could be thermally formed in the early universe and we assume that the stable $d^*$(2380)-BECs constitute all of the dark matter in our universe.  

As the universe expands, the matter temperature would decrease quickly to much less than 1 MeV. Therefore, after the $d^*$(2380)-BECs thermally formed in the early universe, they would cool down quickly and soon become very stable as the collisional kinetic energy is not enough to break down the $d^*$(2380)-BECs. However, after galaxies formed, some exotic astrophysical phenomena (e.g. supernovae) would emit a large amount of high-energy photons and cosmic rays. The energies of these particles can be larger than the energy threshold ($>1$ TeV) such that they can break down the stable $d^*$(2380)-BECs to free $d^*$(2380) hexaquarks. The free $d^*$(2380) hexaquarks are unstable and they will quickly decay to other elementary particles (e.g. gamma-ray photons) without forming back to a $d^*$(2380)-BEC. The energy spectra of the free $d^*$(2380) hexaquarks can be calculated numerically by considering several major decaying channels (e.g. via pion, proton, neutron and deuteron channels) \citep{Bashkanov}.     

Therefore, the decay rate of the $d^*$(2380)-BECs $\Gamma_d$ would be dependent on the amount of the high-energy cosmic rays (including gamma rays), which is proportional to the number density of the high-energy cosmic rays $n_{CR}$. The interaction rate between high-energy cosmic rays and $d^*$(2380)-BECs in a particular volume is given by $n_{CR}N_d \sigma_{CR,d}c$, where $N_d$ is the total number of $d^*$(2380)-BECs inside the volume and $\sigma_{CR,d}$ is the cross section of the interaction. Suppose that the energy of the cosmic rays is greater than the threshold break down energy of a $d^*$(2380)-BEC ($E \sim 1$ TeV). The $d^*$(2380)-BECs would break down to give a large amount of free $d^*$(2380) hexaquarks and they will decay in a very short time ($\sim 10^{-23}$ s). Therefore, the number of decaying $d^*$(2380)-BECs per unit time $N_d \Gamma_d$ is equal to the interaction rate:
\begin{equation}
n_{CR}N_d \sigma_{CR,d}c=N_d \Gamma_d.
\end{equation}
Following the above equation, we get the decay rate $\Gamma_d=n_{CR} \sigma_{CR,d}c$. 

On the other hand, the cosmic-ray flux $\Phi_{CR}$ (in cm$^{-2}$ s$^{-1}$) emitted from a volume is proportional to the number density of the cosmic rays $n_{CR}$ by $\Phi_{CR}=n_{CR}c/4$ (like blackbody radiation emission). Therefore, we get $\Gamma_d=4 \sigma_{CR,d} \Phi_{CR}$. For thousands of $d^*$(2380) hexaquarks forming a $d^*$(2380)-BEC, the size of a $d^*$(2380)-BEC can be as large as $R=aD^{1/3} \sim 10^{-12}$ cm \citep{Bashkanov}, where we have assumed the self-interaction length $a \sim 1$ fm and $D=10^3$. The cross section is approximately given by $\sigma_{CR,d}=\pi R^2$. Hence, we have 
\begin{equation}
\Gamma_d=4 \pi R^2 \Phi_{CR}. 
\end{equation}

It is worth noting that the size of a $d^*$(2380)-BEC is large enough to have non-negligible interactions between the $d^*$(2380)-BECs. If they could have elastic collisions, the geometrical self-interaction cross section is given by $\sigma_{dd}=4 \pi R^2$. For $D=10^3$, the rest mass of a $d^*$(2380)-BEC is $m_{BEC} \sim 1$ TeV. Therefore, the cross section per unit mass is $\sigma_{dd}/m_{BEC} \sim 0.01$ cm$^2$/g. This value is below the current observed upper limits of the self-interacting dark matter model ($\sim 0.1-1$ cm$^2$/g) \citep{Randall,Peter}. Therefore, the proposed size and mass of the $d^*$(2380)-BEC dark matter is consistent with the observed limits. For $D>10^3$, the value of $\sigma_{dd}/m_{BEC}$ would be less than $0.01$ cm$^2$/g. Although the value of $\sigma_{dd}/m_{BEC}$ may not be large enough to form large density cores in galaxies \citep{Chan5,Robles}, the possible interactions between the $d^*$(2380)-BECs or the interactions between the $d^*$(2380)-BECs and baryons might have some other interesting implications that require further explorations. Nevertheless, the above geometrical approach does not include the possible quantum mechanical interactions between the $d^*$(2380)-BECs (e.g. tunneling effect) and between $d^*$(2380)-BECs and baryons. Some suppression or enhancement of interaction cross sections might appear so that the actual interaction cross sections would be quite different from our estimated value. If the enhancement is significant so that $\sigma_{dd}/m_{BEC} \sim 0.1-1$ cm$^2$/g, it may be able to account for the dark matter density cores observed in galaxies. Future experimental investigations on the $d^*$(2380) interactions can give better hints for this issue. On the other hand, since the cosmic rays we considered have a higher energy than the break down energy of a $d^*$(2380)-BEC, the geometrical argument applied in Eq.~(3) is still valid.

Consider our Milky Way Galaxy as an example. The background diffuse TeV cosmic rays in our Galaxy would break down the $d^*$(2380)-BECs to give free $d^*$(2380) hexaquarks. Therefore, we can predict the average decay rate $\Gamma_d$ in our Galaxy by using the background diffuse cosmic-ray data. The isotropic background TeV gamma-ray flux measured is $<10^{-11}$ cm$^{-2}$ s$^{-1}$ sr$^{-1}$ \citep{Ackermann2,Harding}, which is much smaller than the isotropic background diffuse TeV electron and positron flux measured by the DAMPE \citep{Ambrosi} and Fermi-LAT \citep{Abdollahi} ($\approx 10^{-8}$ cm$^{-2}$ s$^{-1}$ sr$^{-1}$). Therefore, the effect of the diffuse background TeV electrons and positrons would be much more dominant in breaking down the $d^*$(2380)-BECs. Using Eq.~(3), the predicted average decay rate of $d^*$(2380)-BECs in our Galaxy is $\Gamma_d \sim 10^{-31}$ s$^{-1}$. Compared with the age of our universe $t \sim 10^{17}$ s, only a small amount of $d^*$(2380)-BECs have decayed. Based on this formulation, the $d^*$(2380)-BECs may be very stable in our universe. However, we expect that the decay rate would vary with different astrophysical environments and it is much larger in a volume surrounding by TeV gamma-ray or cosmic-ray sources.

\section{Constraining the decay rate by astronomical data}
The energy of photons emitted from the decay of a free $d^*$(2380) hexaquark is $E=100-500$ MeV \citep{Bashkanov}. These photons would quickly contribute to the isotropic background gamma-ray spectrum. Therefore, the observational isotropic gamma-ray background data of energy $100-500$ MeV might be able to constrain the decay rate $\Gamma_d$. The isotropic gamma-ray background (IGRB) was well-measured in the past decade \citep{Ackermann2}, which can give stringent constraints for $\Gamma_d$.

The gamma-ray flux $\phi$ (in GeV cm$^{-2}$ s$^{-1}$ sr$^{-1}$) emitted by the decay of the $d^*$(2380) hexaquarks is given by
\begin{equation}
\phi= \frac{\bar{J}}{4 \pi m_d}\Gamma_d \left(E^2 \frac{dN}{dE} \right),
\end{equation}
where $dN/dE$ is the energy spectrum (in GeV$^{-1}$) of the decay per $d^*$(2380) hexaquark and $\bar{J}$ is the J-factor per unit solid angle, which is defined as
\begin{equation}
\bar{J}=\frac{1}{\Delta \Omega} \int \rho ds \int d\Omega. 
\end{equation}
Here, $\rho$ is the density of dark matter (i.e. $d^*$(2380)-BECs), $s$ is the line-of-sight distance and $\Delta \Omega$ is the solid angle.

Recently, a comprehensive analysis considering different dark matter density profiles and different baryonic models has been done for our Milky Way Galaxy \citep{Lin}. It shows that the Navarro-Frenk-White (NFW) dark matter density profile with particular bulge and disk models (B7D1G1 and B6D1G1 models) can give the best fits for the rotation curve data out to $\sim 100$ kpc obtained in \citet{Huang}. Following this best-fit result with uncertainties and taking the distance to the Galactic Center $D_L=8$ kpc, we get $\bar{J}=0.064-0.073$ g cm$^{-2}$ sr$^{-1}$ for the isotropic emission. The $1 \sigma$ upper limit of the observed residual gamma-ray flux within $E=140-200$ MeV is $1.06 \times 10^{-6}$ GeV cm$^{-2}$ s$^{-1}$ sr$^{-1}$ \citep{Ackermann2}. It gives $\Gamma_d=(7.5-8.6) \times 10^{-24}$ s$^{-1}$, which is the upper limit of the average decay rate in our Galaxy. We can see that this upper limit is several orders of magnitude larger than our predicted value $\Gamma_d \sim 10^{-31}$ s$^{-1}$. Therefore, it is very difficult for us to constrain the observed decay rate down to the predicted value based on the current observational data and techniques. 

In Fig.~1, we show the calculated energy spectrum $\phi$ and compare it with the observational data. We can see that there are two peaks in the energy spectrum. However, the peak at a lower energy ($E \approx 180$ MeV) is much more dominant than the one at a higher energy ($E \approx 450$ MeV). Therefore, if one can constrain the decay rate upper limit down to the predicted value, the decaying feature of the $d^*$(2380) hexaquarks can be best verified or falsified by the energy spectrum near $E=180$ MeV.  

\section{Discussion}
The $d^*$(2380)-BEC dark matter model is very attractive. It is because no extra elementary particle beyond the Standard Model or theory beyond the General Relativity is required to account for the dark matter. The existence of $d^*$(2380) hexaquarks has been verified by experiments. Moreover, the $d^*$(2380) hexaquarks can form stable BECs and these BECs can be thermally produced in the early universe \citep{Bashkanov}. These properties make the $d^*$(2380)-BEC a very competitive candidate of dark matter and the entire model is very simple. Although the details of the strong interaction between $d^*$(2380) are not well-understood, future particle experiments might be able to determine these details.

One important potential signature of this dark matter model is the decaying signal of the free $d^*$(2380) hexaquarks. The stable $d^*$(2380)-BECs could be destroyed by the high-energy cosmic rays so that the free $d^*$(2380) hexaquarks would spontaneously produce a large amount of $\sim 100-500$ MeV gamma rays. This process can be characterized by the decay rate $\Gamma_d$. In this article, we show that the value of $\Gamma_d$ is strongly correlated with the high-energy cosmic-ray flux. Generally speaking, the average decay rate $\Gamma_d$ within a volume is larger when there exist a larger flux of TeV cosmic rays. Therefore, the decay rate $\Gamma_d$ may not be a constant, which is different from the predictions of other decaying dark matter models (e.g. decaying sterile neutrinos \citep{Boyarsky}). Nevertheless, one interesting feature of this non-constant decay rate is that if there exist some extremely exotic high-energy astrophysical phenomena in a galaxy, the amount of cosmic rays may be large enough to destroy most of the $d^*$(2380)-BEC dark matter. Then, the galaxy would become a galaxy lacking dark matter, like NGC1052-DF2 and NGC1052-DF4 \citep{vanDokkum,vanDokkum2}.

Beside the theoretical prediction of $\Gamma_d$, we also constrain the $\Gamma_d$ by observational data. In fact, a recent study has performed a similar analysis to constrain the decay rate $\Gamma_d$ by astronomical data of different structures \citep{Beck}. It shows that the Milky Way data can give the tightest constraint for the decay rate ($\Gamma \le 3.9 \times 10^{-24}$ s$^{-1}$). However, the J-factor obtained in that study using the CLUMPY code originates from the assumption of dark matter annihilation but not decay \citep{Hutten}, and the uncertainties of the J-factor is very large (approximately an order of magnitude) \citep{Beck}. In our analysis, the J-factor is calculated using the latest comprehensive study of our Galaxy, which has considered four different dark matter density profiles and 56 combinations of baryonic models. Therefore, our constraints would be more robust and contain less systematic uncertainties. 

However, based on the observational data of the IGRB, we find that the predicted decay rate is much less than our current observed upper limit. This suggests that verifying this dark matter model using decaying signature is very difficult. Such decaying signature might be stronger near the active galactic nuclei or the sources of some exotic astronomical phenomena (e.g. supernovae, black hole mergers). Future observations focusing on $\sim 100-200$ MeV gamma rays might be possible to verify or falsify the $d^*$(2380)-BEC dark matter model, though it would not be an easy task. 

Beside the decaying signature, the size of a $d^*$(2380)-BEC is not negligibly small so that the scattering between $d^*$(2380)-BECs or between $d^*$(2380)-BECs and baryons might be able to provide some observable signatures, such as the formation of small soft cores in dwarf galaxies \citep{Fitts,Kahlhoefer} or the correlation between the dynamical mass and baryonic mass in galaxies and galaxy clusters \citep{Chan6,Chan7}. We have applied a simple geometrical approach to estimate the self-interaction cross section between the $d^*$(2380)-BECs. However, due to our poor understanding of the $d^*$(2380) interactions, the actual value of the cross section might be different from our estimated value. Therefore, it is still possible that the quantum-enhanced self-interaction cross section of $d^*$(2380)-BECs is large enough (e.g. $\sigma_{dd}/m_{BEC} \sim 1$ cm$^2$/g) to form large core structures in galaxies. Extensive investigations in particle experiments, theoretical simulations and astronomical observations are required to verify this interesting dark matter model.

\begin{figure}
 \includegraphics[width=140mm]{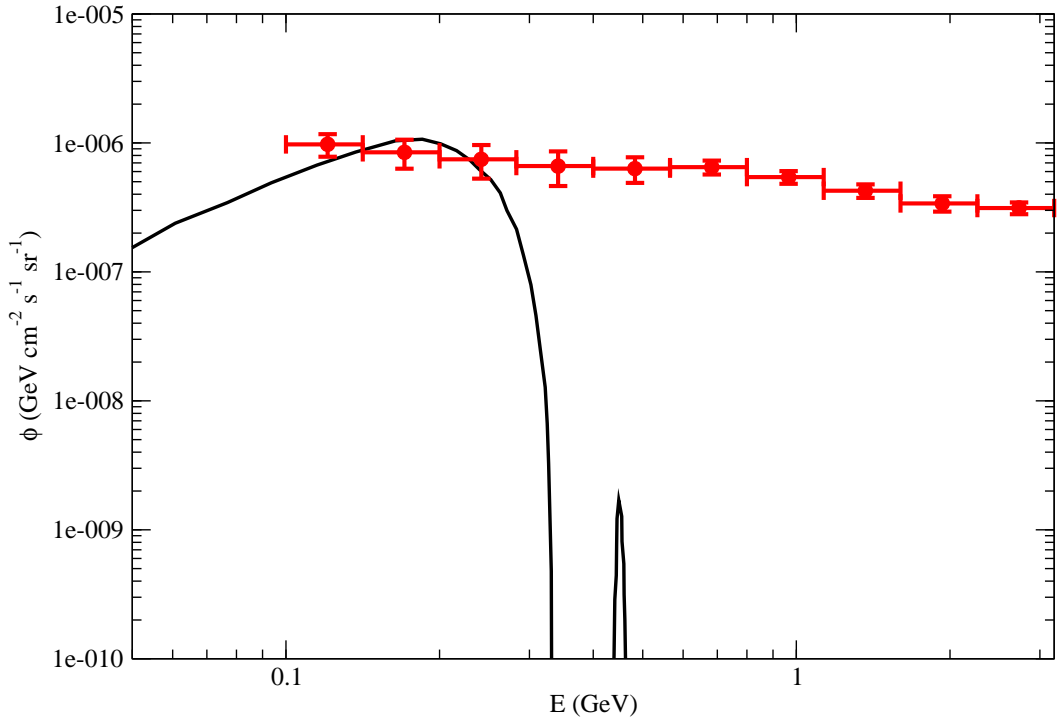}
\caption{The black solid line represents the predicted gamma-ray background spectrum originates from the decay of $d^*$(2380)-BECs in our Galaxy, assuming $\Gamma_d=7.8 \times 10^{-24}$ s$^{-1}$ and $\bar{J}=0.070$ g cm$^{-2}$ sr$^{-1}$. The data with error bars are the isotropic gamma-ray background (IGRB) measured by Fermi-LAT \citep{Ackermann2}.}
\label{Fig1}
\end{figure}

\begin{acknowledgements}
The work described in this paper was supported by a grant from the Research Grants Council of the Hong Kong Special Administrative Region, China (Project No. EdUHK 28300518).
\end{acknowledgements}

\end{document}